\documentclass[aps,pra,preprint,showpacs]{revtex4-1}
\usepackage{graphicx,dcolumn,bm,xcolor,amsmath,soul}

\begin{document}

\title{External light control of three-dimensional ultrashort far-infrared pulses in an inhomogeneous array of carbon nanotubes}
\author{Eduard G. Fedorov} \affiliation{
ITMO University, 197101 Saint Petersburg, Russia}
\author{Alexander V. Zhukov} \affiliation{Singapore University of Technology
  and Design, 8 Somapah Road, 487372 Singapore} \affiliation{Entropique Group
  Ltd., 3 Spylaw Street, Maori Hill, 9010 Dunedin, New Zealand}
\author{Roland Bouffanais} \affiliation{Department of Mechanical Engineering, University of Ottawa, Ottawa ON K1N 6N5, Canada}
\author{Natalia N. Konobeeva} \affiliation{Volgograd State University, 400062, Volgograd, Russia.}
\author{Evgeniya V. Boroznina} \affiliation{Volgograd State University, 
  400062 Volgograd, Russia}
\author{Boris A. Malomed} \affiliation{Department of Physical Electronics,
  School of Electrical Engineering, Faculty of Engineering, Tel Aviv
  University, 69978 Tel Aviv, Israel} \affiliation{Instituto de Alta Investigaci\'{o}n, Universidad de Tarapac\'{a}, Casilla 7D, Arica, Chile.}
\author{Herv{\'e} Leblond} \affiliation{LUNAM Universit{\'e}, Universit{\'e}
  d'Angers, Laboratoire de Photonique d'Angers, EA 4464, 2 Boulevard
  Lavoisier, 49000 Angers, France}
\author{Dumitru Mihalache} \affiliation{Academy of Romanian Scientists, 54
  Splaiul Independentei, Bucharest, RO-050094, Romania} \affiliation{Horia
  Hulubei National Institute of Physics and Nuclear Engineering, Magurele,
  RO-077125, Romania}
\author{Mikhail B. Belonenko} 
\affiliation{Volgograd State University, 400062 Volgograd, Russia}
\affiliation{Entropique Group Ltd., 3 Spylaw Street, Maori Hill, 9010 Dunedin,
  New Zealand}
\author{Nikolay N. Rosanov} \affiliation{Ioffe Institute, 194021 Saint Petersburg, Russia} \affiliation{ITMO University, 197101 Saint Petersburg, Russia}
\author{Thomas F. George} \affiliation{Office of the Chancellor, Departments of
  Chemistry \& Biochemistry and Physics \& Astronomy, University of
  Missouri-St. Louis, St. Louis, Missouri 63121, USA}
\date{\today }

\begin{abstract}
We present a study of the propagation of three-dimensional (3D) bipolar
electromagnetic ultrashort pulses in an inhomogeneous array of semiconductor
carbon nanotubes (CNTs) in the presence of a control high-frequency (HF)
electric field. The inhomogeneity is present in the form of a layer with an
increased concentration of conduction electrons, which acts as a barrier for
the propagation of ultrashort electromagnetic pulses through the CNT array.
The dynamics of the pulse is described by a nonlinear equation for the
vector potential of the electromagnetic field (it takes the form of a 3D
generalization of the sine-Gordon equation), derived from the Maxwell's
equations and averaged over the period of the HF control field. By means of
systematic simulations, we demonstrate that, depending on the amplitude and
frequency of the HF control, the ultrashort pulse approaching the barrier
layer either passes it or bounces back. The layer's transmissivity for the
incident pulse is significantly affected by the amplitude and frequency of
the HF control field, with the reflection coefficient nearly vanishing in
intervals that make up a discrete set of transparency windows, which
resembles the effect of the electromagnetically-induced transparency. Having
passed the barrier, the ultrashort pulse continues to propagate, keeping its
spatiotemporal integrity. The results may be used for the design of soliton
valves, with the transmissivity of the soliton stream accurately controlled
by the HF field.
\end{abstract}

\pacs{42.65.Tg, 42.65.Sf, 78.67.-n, 78.67.Ch}
\maketitle


\section{Introduction}


%
Modern laser technologies offer a variety of opportunities for generating
ultrashort pulses corresponding to several half-periods of field
oscillations~\cite{a,b,1,2}. This has provided the impetus for studies of the
formation and propagation of nonlinear electromagnetic waves in various
media~\cite{c,d,e,f,g,3,4,5}. In this connection, graphene-based materials have
attracted attention as promising media for both basic research and practical
applications in the fields of photonics and optoelectronics (e.g., see
reviews~\cite{6,7} and references therein). In particular, carbon nanotubes
(CNTs)---quasi-one-dimensional carbon macromolecules~\cite{8,9,10}---offer
high potential for the development of optoelectronic devices, based on the
propagation of nonlinear electromagnetic waves, such as ultrafast laser
pulses. These may be photodetectors, solar energy converters, transparent
conductive surfaces, displays, etc.

From the point of view of the potential applications to optoelectronics,
interest in carbon nanotubes is due to the peculiarity of their electronic
structure. In particular, the nonparabolicity of the dispersion of
conduction electrons (the dependence of the energy on the quasimomentum)
leads to a significant nonlinearity in the response of nanotubes to the
application of a moderate electromagnetic field with intensities starting
from $10^{3}$--$10^{4}$ V/cm (see. e.g., Ref.~\cite{11}). This circumstance
makes it possible to observe a number of unique physical phenomena in
nanotube media, including nonlinear diffraction, self-focusing of laser
beams, propagation of solitons, etc.~\cite{12,13,14}.

The possibility of propagation of infrared solitary electromagnetic waves in
arrays of CNTs was first theoretically established in the approximation of a
uniform field along the axis of nanotubes in a one-dimensional (1D) model in
Ref.~\cite{14}. Subsequently, the
possibility of the propagation and interaction of solitary electromagnetic
waves in CNT\ arrays was studied in a 2D model, using the same approximation
as mentioned above, \cite{15,16,17}, and taking into account localization of
the field in directions orthogonal to the propagation of the electromagnetic
wave~\cite{18}. In Refs.~\cite{19,20}, taking into account the most
fundamental 3D spatial localization of the laser pulse field, the
propagation and interaction of solitary electromagnetic waves in nanotube
arrays in the 3D geometry has been studied.

The previous studies have established that the evolution of electromagnetic
waves in arrays of semiconductor CNTs substantially depends on various
physical factors, such as the presence of various impurities, as well as
static and dynamic inhomogeneities. In particular, doping a sample with a
uniformly distributed multilevel impurity can lead to a modification of the
parameters of a propagating electromagnetic pulse, as compared to the
propagation in pure samples~\cite{21}. In addition, dynamic inhomogeneities
of the spatial distribution of the concentration of conduction electrons in
CNT\ arrays induced by laser pulses can serve as mediators in the
interaction of extremely short pulses~\cite{20}. Besides that, the
interaction of ultrashort pulses with static localized inhomogeneities in
the array deserves special attention from the perspective of possible
applications (see, e.g., Ref.~\cite{22}). For example, as a result of the
interaction of the ultrashort pulses with a layer carrying high electron
density (HED), selective nature of the pulse scattering by such a layer has
been established, offering new possibilities for developing light control
methods in micro- and nanostructures~\cite{23,24,25}.

The concept of controlling the dynamics of ultrashort pulses, proposed in
recent works, suggests various outcomes of the interaction of a solitary
wave with the structural inhomogeneity of the medium, depending on both
parameters of the pulse itself and properties of the inhomogeneity. For
example, a pulse with an amplitude significantly exceeding a certain
threshold value can pass the HED layer, while a pulse with the amplitude
falling below the threshold will be reflected from the layer. In this case,
decrease in the thickness of the inhomogeneity layer also facilitates the
passage of the pulse through such the layer. However, properties of media
acting as the waveguides are usually fixed by the manufacturing procedure.
Parameters of the laser pulses cannot be easily adjusted either if the puse
stream is generated by a standard source. Therefore, possibilities for the
design of the control of the pulse dynamics in micro- and nanostructures are
limited, and it is relevant to develop a method for controlling the dynamics
of ultrashort pulses in inhomogeneous nonlinear media by means of an
independent control tool. In this work we demonstrate that externally
applied high-frequency (HF) electric field may provide such a tool, which
acts by dynamically modifying properties of the electronic subsystem in the
CNT array. The control HF field is switched on for a time 
significantly exceeding the characteristic duration of the extremely short 
pulse scattered by the HED layer. During the presence of the control HF 
field, the pulse has the time to enter the system, to adjust its parameters to 
the properties of the medium---modified by this control HF field---and to interact 
(i.e., to perform the act of scattering/collision, transmission or reflection) with 
the HED layer. The results is the creation of transparency windows at specific
values of the amplitude of the control HF field, at which the reflection
coefficient practically vanishes. This effect is similar to the well-known
phenomenon of the electromagnetically-induced transparency~\cite{EIT0,EIT,EIT2}. Further, turning the external field on/off may allow
ultrafast switching of the interaction of ultrashort pulses with the HED
layer, which thus acts as a controllable semi-transparent mirror for the
pulses. As a result, pulses with the same parameters can either pass the
layer or bounce back from it, depending on the presence of the HF control
field. Thus, in this work we address the dynamics of 3D ultrashort
electromagnetic pulses in the bulk array of semiconductor CNTs, with the HED
layer embedded into it, which acts as a controllable obstacle for the
transmission of pulses through the system.

The rest of the paper is organized as follows. The system is formulated in
Section II, the basic evolution equation for the electromagnetic field
carrying the ultrashort pulse is derived in Section III, and characteristics
of the field are presented in Section IV. Numerical results for the
transmission of the pulse in the CNT array and its interaction with the HED
layer, in the absence and presence of the HF control field, are
systematically reported in Section V, where, in particular, the existence of
the above-mentioned transparency windows is demonstrated. Finally, main
findings of the work are summarized in Section VI.

\section{The system}

\begin{figure}[tbp]
\includegraphics[width=0.6\textwidth]{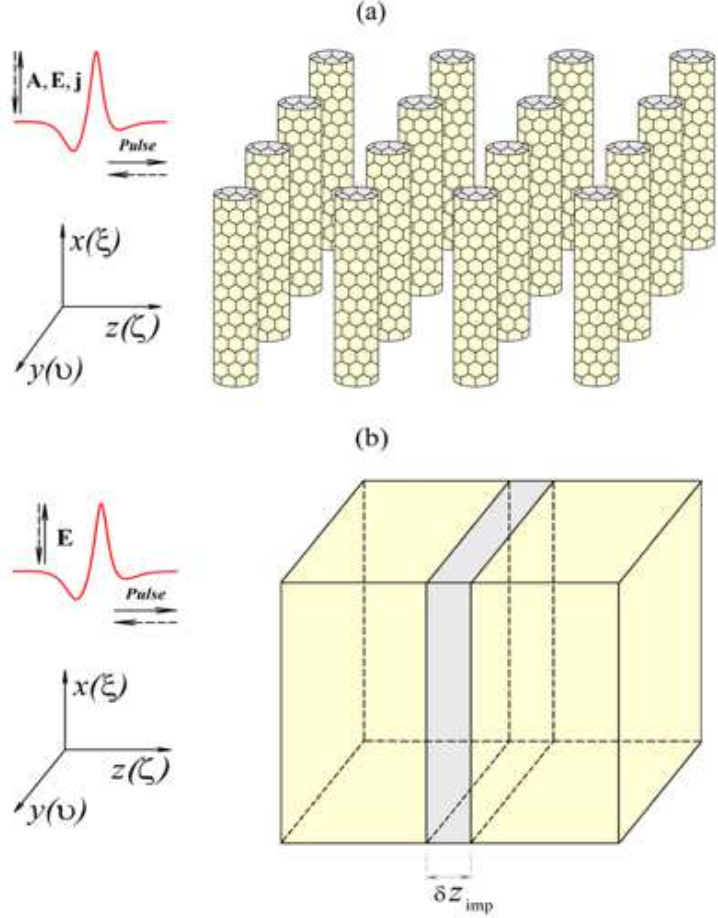}
\caption{Geometry of the system: (a) Schematic representation of an array of
nanotubes and the orientation of vectors of physical quantities in the
Cartesian coordinate system. (b) Location of a layer of increased electron
concentration.}
\label{fig1}
\end{figure}

%
We consider the propagation of a bipolar solitary electromagnetic wave in the bulk
array of single-walled semiconductor CNTs, embedded in an inhomogeneous
dielectric medium, under the influence of an external HF control
electromagnetic field. The vector of the electric field of a bipolar pulse should have opposite directions at different instants of time at a fixed point. It is assumed that space between the CNTs is filled with a dielectric, while the heterogeneity of the array is represented by a layer with an increased concentration of conduction electrons. The CNTs considered here are of the \textquotedblleft zigzag" type $(m,0)$, where integer $m$ (different from a multiple of three for semiconductor nanotubes) determines the CNT radius, $R=mb\sqrt{3}/2\pi $, where $b=1.42\times 10^{-8}~\mathrm{~cm}$ is the distance between adjacent carbon atoms~\cite{8,9,10}.
The CNTs are arranged in such a way that their axes are parallel to the
common $x$-axis, and an ultrashort laser pulse propagates along the $z$%
-axis, that is, in a direction perpendicular to CNT axes. In this case, the
electric field of the pulse, $\mathbf{E}=\{E,0,0\}$, is collinear to the $x$%
-axis, see Fig.~\ref{fig1}. We assume that the overlap of wave functions of
electrons between adjacent CNTs is negligible (no tunneling of electrons
between adjacent nanotubes), and the system under consideration is
electrically quasi-one-dimensional, featuring conductivity only along the $x$%
-axis.

Further, we assume that, in addition to the field of the ultrashort
electromagnetic pulse, the system includes an external control HF electric
field $\mathbf{E}_{1}=\{E_{1},0,0\}$, with $E_{1}=E_{10}\cos (\omega
_{1}t+\alpha )$, where $E_{10}$, $\omega _{1}$, and $\alpha $ are the
amplitude, frequency and phase shift of the control field. We also assume
that frequency $\omega _{1}$ significantly exceeds the inverse of the
characteristic pulse's temporal width: $\omega _{1}\gg 2\pi /\Delta t_{\text{%
pulse}}$. HF field $\mathbf{E}_{1}$ may be realized by exciting transverse
standing electromagnetic wave modes of the embedding waveguide, by means of an appropriate laser source.

The CNT array, which is a discrete structure at the microscopic level, is
considered in this paper in the approximation of a continuum medium, in the
context of the interaction with the electromagnetic pulse and the external
HF radiation (the control field). This approximation is valid for a wide
range of system's parameters---in particular, when the wavelength of the
external HF radiation and the characteristic distance of the variation of
the pulse's field substantially exceed the separation between adjacent CNTs,
as well as the mean free-path of conduction electrons along the CNT axis.
For example, the CNT radius $R\approx 5.5\times 10^{-8}\mathrm{~cm}$ and $%
m=7 $ produce the separation between them (sufficient to ensure the
conductivity of in the CNT array only along the $x$-axis)---even if
substantially exceeding radius $R$---as negligibly small in comparison with
wavelengths of the electromagnetic radiation in the infrared range.

Given the orientation of the coordinate system axes relative to the nanotube
axis chosen in Fig.~\ref{fig1}, the electron energy spectrum for CNTs takes
the form of 
\begin{equation}
\epsilon (p_{x},s)=\gamma _{0}\sqrt{1+4\cos \left( p_{x}\frac{d_{x}}{\hbar }%
\right) \cos \left( \pi \frac{s}{m}\right) +4\cos ^{2}\left( \pi \frac{s}{m}%
\right) },  \label{1}
\end{equation}%
where $\gamma_0 = 2.7$ eV, the electron quasimomentum is $\mathbf{p}=\left\{ p_{x},s\right\} $, $%
p_{x}$ is the projection of the quasimomentum of the conduction electron
onto the CNT\ axis, and $s$ is an integer characterizing the momentum
quantization along the perimeter of the nanotube, $s=1,2,\dots ,m$. Here $m$
is the number of hexagonal carbon cycles forming the circumference of a
nanotube, $\gamma _{0}$ is the overlap integral, and $d_{x}=3b/2$~\cite%
{8,9,10}. 

We solve the problem in the semiclassical approximation, thus requiring the
following conditions to be satisfied: $\hbar \omega _{0}\ll 2\gamma _{0}$ ($%
\omega _{0}$ is the characteristic frequency of the electronic subsystem of
CNTs), $|Ee|d\ll 2\gamma _{0}$, $|E_{1}e|d\ll 2\gamma _{0}$, $\hbar /\Delta
t_{\text{pulse}}\ll 2\gamma _{0}$. We adopt one more important assumption
regarding the ratio of the duration of the electromagnetic pulse, $\Delta t_{%
\text{pulse}}$, relaxation time $t_{\text{rel}}$ of the conduction current
along the axis of the nanotubes, and the length of the time interval $\Delta
t$ for observing the evolution of the electromagnetic field in the system.
Specifically, we assume that the observation time substantially exceeds the
characteristic pulse duration, but is still shorter than the relaxation
time: $\Delta t_{\text{pulse}}\ll \Delta t<t_{\text{rel}}$. This condition
allows us to maintain the collisionless approximation, in which we can
neglect the influence of collisions of electrons with defects of the CNT
array on the evolution of the conduction current and pulse's electromagnetic
field. It should be noted that typical scattering times are strongly 
temperature dependent \cite{JDD}, as $t_{\text{rel}} \propto\sinh (1/T)$. 
Thus, the question of collisionless approximation applicability is only about 
cooling the sample during the experiment correspondingly.

\section{The evolution equation for the electromagnetic field of short
pulses}

\subsection{Density of conductivity electrons}


Starting from the full system of the Maxwell's equations~\cite{26,27}, we
derive the following wave-propagation equation, in the geometry under the
consideration:
\begin{equation}
\frac{\varepsilon }{c^{2}}\frac{\partial ^{2}A}{\partial t^{2}}-\frac{%
\partial ^{2}A}{\partial y^{2}}-\frac{\partial ^{2}A}{\partial z^{2}}-\frac{%
4\pi }{c}j=0,  \label{2}
\end{equation}%
where $A(x,y,z,t)$ and $j(x,y,z,t)$ are the projections of the vector
potential $\mathbf{A}=\{A,0,0\}$ and current density $\mathbf{j}=\{j,0,0\}$
onto the $x$-axis, and $c$ is the speed of light in vacuum. The electric
field of the laser pulse is then $\mathbf{E}=-c^{-1}\partial \mathbf{A}%
/\partial t$~\cite{26,27}.

The nonuniformity (localization) of the field along the $x$-axis drives
evolution and spatial nonuniformity of the electron concentration in the
sample, due to the action of the conductivity along the axis of the
nanotubes, while the field nonuniformity along directions orthogonal to the
axes of the nanotubes does not contribute to the redistribution of the
electron concentration, due to negligible overlap of the wave functions of
the electrons in adjacent nanotubes and the absence of conductivity in the $%
\left( y,z\right) $ plane. Full analysis of the accumulation of electric
charge and, accordingly, taking into account the field of this charge is a
separate problem that is beyond the scope of this work. However, as shown by
numerical simulations performed earlier (see Refs.~\cite{18,19,20}),
differences in electron concentration (dynamic inhomogeneities) emerging in
the course of the passage of electromagnetic pulses in the sample have the
magnitude of few percent relative to the initial equilibrium concentration, $%
n_{0}$. In this case, there is no significant disturbance in the dynamics of
the pulses with respect to the results obtained in the framework of the
approximating admitting uniform field along the CNT\ axis (see, e.g., Ref.~%
\cite{28}). Thus, when considering ultrashort electromagnetic pulses,
subject to the above-mentioned condition $\Delta t_{\text{pulse}}\ll t_{%
\text{rel}}$, the nonstationary disturbance in the concentration of
conduction electrons may be neglected.

Based on these considerations, we assume that the distribution of the
concentration of conduction electrons in the sample remains approximately
constant, in accordance with the approximation of the uniform electric field
acting in the axial direction. Thus, equations for the concentration of
conduction electrons and scalar potential may be excluded from the system
under the consideration. As a result, for the propagation of the short
electromagnetic pulse through the CNT array, the evolution of the field in
the array is described, with reasonable accuracy, by the single equation~%
\eqref{2} for the vector potential.

The projection of the conduction current density $j$ onto the CNT\ axis is
determined using the approach developed in Refs.~\cite{29,30}, which yields
\begin{equation}
j=2e\sum_{s=1}^{m}\int\limits_{-\pi \hbar /d}^{+\pi \hbar
/d}v_{x}f(p_{x},s)dp_{x},  \label{3}
\end{equation}%
where $e$ is the electron charge ($e<0$), $v_{x}$ and $f(p_{x},s)$ are the
electron velocity and distribution function over quasi-momenta $p_{x}$, and
numbers $s$ characterize, as said above, the quantization of the electron
momentum along the perimeter of the nanotube. Factor $2$ in Eq.~\eqref{3}
takes into account the summation of electrons over spins, and the
integration over the quasimomentum is carried out within the first Brillouin
zone. Using the expression for the energy of electrons \eqref{1} in
determining their velocity $v_{x}=\partial \epsilon (p_{x},s)/\partial p_{x}$,
and taking into account the Fermi-Dirac distribution of electrons according
to Eq. \eqref{3}, we obtain an expression for the current density (further
details of the derivation can be found in recent work~\cite{20}):
\begin{equation}
j=-en\frac{d_{x}}{\hbar }\gamma _{0}\sum_{r=1}^{\infty }G_{r}\sin \left( rA%
\frac{ed_{x}}{c\hbar }\right) .  \label{4}
\end{equation}%
Here, $n$ is the concentration of conduction electrons at a given point in
the volume of the sample, and coefficients $G_{r}$ are determined as
\begin{equation}
G_{r}=-r\frac{\sum_{s=1}^{m}\frac{\delta _{r,s}}{\gamma _{0}}\int_{-\pi
}^{+\pi }\cos (r\theta)\left\{ 1+\exp \left[ \frac{\theta _{0,s}}{2}%
+\sum_{q=1}^{m}\theta _{q,x}\cos (q\theta )\right] \right\} ^{-1}d\theta }{%
\sum_{s=1}^{m}\int_{-\pi }^{+\pi }\left\{ 1+\exp \left[ \frac{\theta _{0,s}}{%
2}+\sum_{q=1}^{m}\theta _{q,x}\cos (q\theta )\right] \right\} ^{-1}d\theta },
\label{5}
\end{equation}%
where $\theta _{r,s}=\delta _{r,s}(k_{B}T)^{-1}$, $k_{B}$ is the Boltzmann
constant, $T$ is temperature, and $\delta _{r,s}$ are coefficients of the
expansion of electron energy \eqref{1} in the Fourier series~\cite{31},
\begin{equation}
\delta _{r,s}=\frac{d_{x}}{\pi \hbar }\int\limits_{-\pi \hbar /d}^{+\pi
\hbar /d}\epsilon (p_{x},s)\cos \left( r\frac{d_{x}}{\hbar }p_{x}\right)
dp_{x}.  \label{6}
\end{equation}

\subsection{Influence the control HF (high-frequency) electric field}

The presence of the field of the ultrashort pulse and external (control) HF
electric field can be taken into account by replacing $A\rightarrow A+A_{1}$
in the expression for current density \eqref{4}. Taking into regard the
definition of the electric field, $\mathbf{E}_{1}=\{E_{1},0,0\}=-c^{-1}%
\partial \mathbf{A}_{1}/\partial t$, this replacement amounts to
\begin{equation}
A\rightarrow A-E_{10}\frac{c}{\omega _{1}}\sin (\omega _{1}t+\alpha ).
\label{7}
\end{equation}%
Further, substituting Eq.~\eqref{7} into expression \eqref{4} for the
current density, and averaging the result over period $2\pi /\omega _{1}$ of
the control HF electric field, we obtain an effective expression for the
current density arising in the sample under the combined action of both the
short-pulse and control fields:
\begin{equation}
\langle j \rangle=-en\frac{d_{x}}{\hbar }\gamma _{0}\sum_{r=1}^{\infty }J_{0}\left( r\frac{%
|eE_{10}|d_{x}}{\hbar \omega _{1}}\right) G_{r}\sin \left( rA\frac{ed_{x}}{%
c\hbar }\right) ,  \label{8}
\end{equation}%
where $J_{0}$ is the zeroth-order Bessel function~\cite{31} (see Appendix A for full details).


\subsection{The HED (high-electron-density) layer}

An increase in the concentration of conduction electrons in a particular
layer can be achieved, for example, by introducing donor dopants at the
stage of fabrication of the sample (a detailed discussion of technical
aspects of doping the CNT array by donors is beyond the scope of this
theoretical paper). We stress that each segment of the sample is assumed to
be electro-neutral; in particular, in the HED layer, the larger charge
density of free electrons is compensated by a balancing higher concentration
of ionized dopants.

We assume that the HED layer is a region of thickness $\delta z_{\text{imp}}$, placed parallel to the CNT\ axes and perpendicular to the axis along which
the ultrashort electromagnetic pulse propagates, see Fig.~\ref{fig1}. We
model the profile of the electron concentration in the sample by a natural
Gaussian, cf. Ref.~\cite{24}:
\begin{equation}
n(z)=n^{\text{bias}}+\left( n_{\text{imp}}^{\text{\text{max}}}-n^{\text{bias}%
}\right) \exp \left\{ -\left( \frac{z}{\delta z_{\text{imp}}}\right)
^{2}\right\} ,  \label{9}
\end{equation}%
where $n_{\text{imp}}^{\text{\text{max}}}$ is the maximum concentration of
conduction electrons in the layer, and $\delta z_{\text{imp}}$ is its
half-width. The concentration of conduction electrons is assumed constant in
any part of the $\left( x,y\right) $ plane.


\subsection{The effective equation for the vector potential}

Substituting the expression for the conduction current density \eqref{8}
into Eq.~\eqref{2}, and taking into account the electron concentration
profile \eqref{9}, we obtain the following effective equation for the
evolution of the vector potential of the ultrashort pulse propagating
through the CNT array, under the action of the control (external) HF field:
\begin{equation}
\frac{\partial ^{2}\Psi }{\partial \tau ^{2}}-\frac{\partial ^{2}\Psi }{%
\partial \xi ^{2}}-\frac{\partial ^{2}\Psi }{\partial \upsilon ^{2}}-\frac{%
\partial ^{2}\Psi }{\partial \zeta ^{2}}+\eta (\zeta )\sum_{r=1}^{\infty
}J_{0}(\kappa r)G_{r}\sin (r\Psi )=0.  \label{10}
\end{equation}%
The notation used in Eq.~\eqref{10} is: $\Psi =\left( ed_{x}/c\hbar \right)
A $ is the dimensionless projection of the vector potential of the
ultrashort pulse onto the CNT\ axis;
\begin{equation}
\tau =\omega _{0}t/\sqrt{\varepsilon },\xi =x\omega _{0}/c,\upsilon =y\omega
_{0}/c,\zeta =z\omega _{0}/c  \label{tau-xi}
\end{equation}%
are dimensionless time and spatial coordinates; $\varepsilon $ is the
averaged relative dielectric constant of the sample (for further details,
see Ref.~\cite{32}); $\eta (\zeta )=n/n^{\text{bias}}$ is the reduced
distribution of the concentration of conduction electrons in the sample,
calculated as per Eq.~\eqref{9}; and coefficients $G_{r}$ are given by
dimensionless expressions \eqref{5} that decrease with the increase of $r$.
Further, the quantity
\begin{equation}
\kappa =\frac{|eE_{10}|d_{x}}{\hbar \omega _{1}},\label{eq:kappa}
\end{equation}  
characterizes the
control HF electric field, and $\omega _{0}=2|e|d_{x}\hbar ^{-1}\left( \pi
n^{\text{bias}}\gamma _{0}\right) ^{1/2}$ is a characteristic frequency of
the electronic subsystem of CNTs. With parameters used in the paper (see
below), it is $\approx \omega _{0}=7.14\times 10^{12}\mathrm{s^{-1}}$,
corresponding to the vacuum wavelength $\lambda _{0}=264~\mathrm{\mu m}$,
which belongs to the far-infrared domain. The central frequency of the waves
under consideration, typically measured by $1/\Delta t_{\text{pulse}}$,
belong to the same range. As a consequence, high frequency $\omega _{1}$ may
be chosen in the mid-infrared domain or even at the highest-frequency edge
of the far-infrared band. Thus, Eq.~\eqref{10} describes the evolution of
the self-consistent field of an the ultrashort pulse interacting with the
electronic subsystem of the CNT array, driven by control HF electromagnetic
field.

\section{Characteristics of the short-pulse's field}

For illustrating the localization of the electromagnetic pulse in space, we
will use the normalized  electromagnetic energy density of the wave, 
$E^{2}\equiv W(\xi ,\upsilon ,\zeta ,\tau )$. Taking into account the well-known formula $\mathbf{E}=-c^{-1}\partial
\mathbf{A}/\partial t$, the energy density 
 of the field selected above can be represented as
\begin{equation}
W=W_{0}\left( \frac{\partial \Psi }{\partial \tau }\right) ^{2},  \label{11}
\end{equation}%
\begin{equation}
W_{0}=E_{0}^{2},E_{0}\equiv -\hbar \omega _{0}/(ed_{x}\sqrt{\varepsilon }).
\label{I0}
\end{equation}%

When the electromagnetic pulse interacts with the HED layer, in the general
case the incident pulse splits in transmitted and reflected waver packets.
The ratio of their energies depends on various factors, including
characteristics of the incident pulse, as well as parameters of the layer
with an increased electron concentration \cite{23,24,25}. As characteristics
of the result of the interaction of the ultrashort pulse with the layer, we
define transmission and reflection coefficients, following Refs.~\cite{24,25}%
:
\begin{align}
K_{\text{pass}}& =\frac{\int_{0}^{+\infty }d\zeta \int_{-\infty }^{+\infty
}d\upsilon \int_{-\infty }^{+\infty }d\xi W(\xi ,\upsilon ,\zeta ,\tau
_{\infty })}{\int_{-\infty }^{+\infty }d\zeta \int_{-\infty }^{+\infty
}d\upsilon \int_{-\infty }^{+\infty }d\xi W(\xi ,\upsilon ,\zeta ,\tau
_{\infty })},  \label{12} \\
K_{\text{refl}}& =\frac{\int_{-\infty }^{0}d\zeta \int_{-\infty }^{+\infty
}d\upsilon \int_{-\infty }^{+\infty }d\xi W(\xi ,\upsilon ,\zeta ,\tau
_{\infty })}{\int_{-\infty }^{+\infty }d\zeta \int_{-\infty }^{+\infty
}d\upsilon \int_{-\infty }^{+\infty }d\xi W(\xi ,\upsilon ,\zeta ,\tau
_{\infty })},  \label{13}
\end{align}%
where $\tau _{\infty }$ corresponds to any instant of time taken after
completion of the interaction of the pulse with the layer, when the pulse is
already located at a sufficiently large distance from the layer, and the
field's energy density at the location of the layer is negligible in comparison
to the maximum pulse's energy density. Coefficients $K_{\text{pass}}$ and $K_{%
\text{refl}}$, defined by Eqs.~\eqref{12} and \eqref{13} represent,
severally, the ratio of the energy of the transmitted and reflected wave
packets to the total field energy in the entire calculation region. The
system considered in this paper being conservative, the coefficients are
subject to relation $K_{\text{refl}}=1-K_{\text{pass}}$, which is a consequence of the energy conservation~\cite{40}. 

If the energy of the wave packet passing through the layer of increased electron concentration substantially exceeds the energy of the reflected wave packet ($K_{\text{pass}} \gg K_{\text{refl}}$), then the electromagnetic pulse passes through this layer. Otherwise, when the energy of the reflected wave packet substantially prevails over the energy of the transmitted wave packet ($K_{\text{pass}} \ll K_{\text{refl}}$), we agree to talk about the reflection of an electromagnetic pulse. At certain values of the system parameters, specifically, at the ``threshold'' value of the initial velocity of the incident pulse, it can be divided into two wave packets with approximately equal energies ($K_{\text{pass}} \approx K_{\text{refl}}$), which after interacting with a layer of increased electron concentration, propagate in opposite directions.

\section{Numerical results}

\subsection{Initial conditions: The shape of the ultrashort electromagnetic
pulse}
Equation \eqref{10} for the vector potential, which governs the evolution of
the field of the ultrashort pulse in the inhomogeneous CNT\ array under the
action of an control HF electric field, is a 3D generalization of the
sine-Gordon equation. As it does not admit exact analytical solutions, we
carried out numerical simulations, taking into account the electron
density-distribution profile \eqref{9}. Following the approach elaborated in
previous works (see Ref.~\cite{25} and references therein), for the initial
condition we take a product of the \textquotedblleft snapshot" of the
dimensionless projection $\Psi _{\Vert }(\zeta ,\tau _{0})$ of the field's
vector potential onto the $\xi $-axis of the CNTs at fixed time moment, $%
\tau =\tau _{0}$, and the initial distribution of the field in the $\left(
\xi ,\upsilon \right) $ plane, orthogonal to the propagation direction of
the pulse:
\begin{equation}
\Psi (\xi ,\upsilon ,\zeta ,\tau _{0})=\Psi _{\Vert }(\zeta )\Psi _{\perp
}(\xi ,\upsilon ).  \label{14}
\end{equation}%
We select the profile $\Psi _{\Vert }(\zeta ,\tau _{0})$ corresponding to
the commonly known breather solution of the sine-Gordon equation \cite{33},
\begin{equation}
\Psi _{\Vert }(\zeta ,\tau )=4\arctan \left\{ \left( \frac{1}{\Omega ^{2}}%
-1\right) ^{1/2}\frac{\sin \chi }{\cosh \mu }\right\} ,  \label{15}
\end{equation}%
where
\begin{align}
\chi & =\sigma \Omega \frac{\tau (\zeta -\zeta _{0})U}{\sqrt{1-U^{2}}},
\label{16} \\
\mu & =\sigma \left\{ \tau U-(\zeta -\zeta _{0})\right\} \sqrt{\frac{%
1-\Omega ^{2}}{1-U^{2}}},  \label{17}
\end{align}%
where $U=u/v$ is the ratio of the initial propagation velocity \eqref{16} of
the pulse along the $\zeta $-axis to the speed of light in the medium, $v=c/%
\sqrt{\varepsilon }$; $\zeta _{0}$ is the dimensionless coordinate of the
center of mass of the pulse along the $\zeta $-axis at the time moment $\tau
=\tau _{0}$; $\Omega \equiv \omega _{B}/\omega _{0}$, with self-oscillation
frequency of the breather $\omega _{B}$ ($0<\Omega <1$); $\sigma \equiv
\sqrt{G_{1}}$; and coefficients $G_{r}$ are calculated using Eq.~\eqref{5}.

A Gaussian was chosen as the transverse profile of the pulse's field, which
is adequate in many settings similar to the present one~\cite{34,35,36}:
\begin{equation}
\Psi _{\perp }(\xi ,\upsilon )=\exp \left( -\frac{\xi ^{2}+\upsilon ^{2}}{%
w_{0}^{2}}\right) ,  \label{18}
\end{equation}%
where $w_{0}$ is the initial transverse size of the pulse at $\tau =\tau
_{0} $. Taking into regard Eqs.~\eqref{15}--\eqref{18}, the projection of
the electric field strength of the pulse onto the CNT axis at $\tau =\tau
_{0}$ is
\begin{equation}
E_{x}=E_{\mathrm{{\text{\text{max}}}}}\frac{\cos \chi \cosh \mu -U\left(
\Omega ^{-2}-1\right) ^{1/2}\sin \chi \sinh \mu }{\cosh ^{2}\mu +\left(
\Omega ^{-2}-1\right) \sin ^{2}\chi }\exp \left( -\frac{\xi ^{2}+\upsilon
^{2}}{w_{0}^{2}}\right) ,  \label{19}
\end{equation}%
where $E_{\mathrm{{\text{\text{max}}}}}=4E_{0}\sigma \sqrt{1-\Omega ^{2}}/%
\sqrt{1-U^{2}}$.

The shape of the electromagnetic pulse generated by initial conditions %
\eqref{15}--\eqref{19} periodically changes. It is called \textit{bipolar}
because electric field \eqref{19} takes both positive and negative values.
The characteristic duration of the pulse can be estimated as (cf. Ref.~\cite%
{25})
\begin{equation}
\Delta t_{\text{pulse}}=2\frac{\ln \left( 2+\sqrt{3}\right) }{\sigma \omega
_{0}}\frac{\sqrt{\varepsilon }}{U}\frac{\sqrt{1-U^{2}}}{\sqrt{1-\Omega ^{2}}}%
.  \label{20}
\end{equation}


\subsection{System's parameters}

As an environment for modeling the propagation of the ultrashort
electromagnetic pulse, we recall that we choose the CNT array of the zigzag
type $(m,0)$: $m=7$, $\gamma _{0}=2.7$ eV, $b=1.42\times 10^{-8}$ cm, $%
d_{x}\approx 2.13\times 10^{-8}$ cm, $n^{\text{bias}}=10^{16}$ cm$^{-3}$, at
temperature $T=293$~K. We assume that the array is embedded in a dielectric
matrix with the effective dielectric constant $\varepsilon =4$ (see Ref.~%
\cite{25} and references therein).

Dimensionless parameter $U$ [see Eqs.~\eqref{16} and \eqref{17}] is varied
within interval $U\in (0.5;0.999)$. Extreme values $U>0.999$ were not
considered due to limitations imposed by the numerical scheme. On the other
hand, at $U<0.5$, the longitudinal width of the pulse begins to approach the
value of the distance traveled by the pulse over a duration $\sim t_{\mathrm{%
rel}}$, which is of no significant practical interest. 

The dimensionless frequency $\Omega $ of internal oscillations of the
initial pulse~\eqref{15} was varied in interval $\Omega \in (0.1;0.9)$. As $%
\Omega $ decreases, the width of the pulse along the the $\zeta $-axis
decreases too, the shape variation being insignificant at $\Omega <0.5$. For
$\Omega >0.9$, the width becomes comparable to the distance traveled by the pulse over time $\sim t_{\mathrm{rel}}$. Transverse width of the pulse was varied in the range
of $1.0\leq w_{0}\leq 10.0$.

When modeling the profile of the electron concentration in the sample, we
varied values of parameter $\eta _{\text{imp}}^{\text{\text{max}}}\equiv n_{%
\text{imp}}^{\text{\text{max}}}/n^{\text{bias}}$ that determines the maximum
electron concentration in the inhomogeneity layer, see Eq.~\eqref{9}, in the
range of $1\leq \eta _{\text{imp}}^{\text{\text{max}}}\leq 100$. The
dimensionless thickness of the HED layer, $\delta \zeta _{\text{imp}}=\delta
z_{\text{imp}}\omega _{0}/c$, was varied from $0.05$ to $0.5$. We stress
that the use of the collisionless approximation (which makes it possible to
consider the system as a conservative one) is justified if the evolution
time is limited by the relaxation time $t_{\text{rel}}$. In particular, with
$t_{\text{rel}}\approx 10$ ps, the ultrashort electromagnetic pulse travels
distance $z\leq ct_{\text{rel}}/\sqrt{\varepsilon }\approx 0.15$ cm in the
medium under the consideration.


\subsection{Interaction of the ultrashort pulse with the HED\
(high-electron-density) layer}

To solve equation \eqref{10} numerically with initial conditions \eqref{15}--%
\eqref{19}, we used an explicit finite-difference three-layer cross-type
scheme for hyperbolic equations described in Refs.~\cite{38,39} and
adapted for the 3D model using the approach reported in detail for the 2D
setting in Ref.~\cite{24}. The calculations produced the electromagnetic
field, $\Psi =\Psi (\xi ,\upsilon ,\zeta ,\tau )$, and the respective field energy density was found as per Eq.~\eqref{11}.

It was thus found that, depending on values of the system's parameters,
various scenarios of the interaction of the ultrashort electromagnetic pulse
with the HED layer are possible, leading to the passage of the layer or
reflection from it. The outcome is determined by control parameters, which
are characteristics of the electromagnetic pulse itself (such as the speed
the incident pulse) and the layer (its thickness and the peak concentration
of conduction electrons in it). The passage of the layer by the pulse is,
naturally, enhanced by both the increase of the incidence speed and the decrease of
the layer's thickness and electron concentration in it, cf. Ref.~\cite{25}
and references therein. The reflection of short pulses
from the HED has a simple explanation. With an increase in the concentration
of carriers in the impurity band, the current induced by the incident pulse
increases too. Thus, the impurity region becomes more conductive, in
comparison to the homogeneous sample, and, 
as a consequence, stronger reflects the electromagnetic wave \cite{25}. On the other
hand, a faster moving pulse has higher energy, which makes it is easier for
it to overcome the effective potential barrier created by the HED layer~\cite{33}.

Parameters of the waveguide medium are fixed by the manufacturing
technology, therefore they cannot be changed to control the pulse-layer
interaction. Parameters of the pulse may be altered, but applications may
make it necessary to control the behavior of pulses with fixed parameters,
created by standard sources. In this context, an essential option, developed
in this work, is to change outcomes of the interaction by means of the
control HF field, i.e., its amplitude and frequency may be used as efficient
control parameters. In fact, the control effect may be achieved not only by
varying these parameters, but also by turning the external field on/off. We
present here results of modeling the propagation of short electromagnetic
pulses in the inhomogeneous CNT\ array, for fixed parameters of the pulse
and HED layer, both in the absence of presence of the control HF field.

\begin{figure}[tbp]
\includegraphics[width=1.0\textwidth]{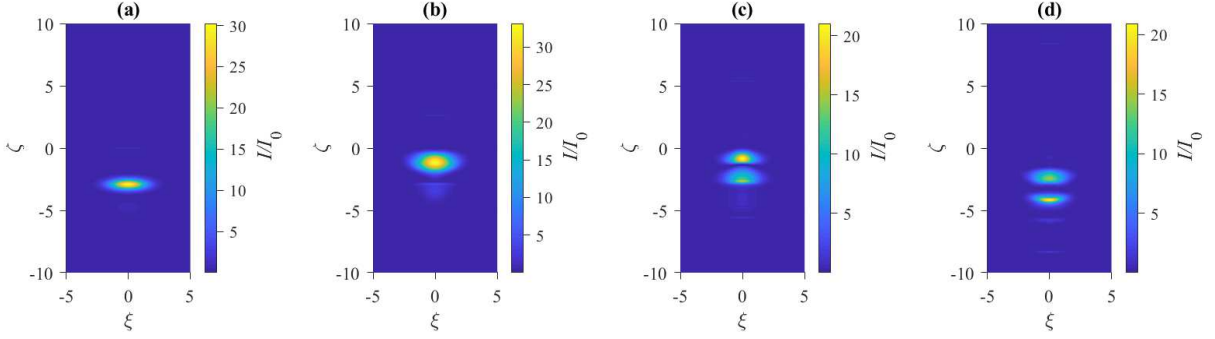}
\caption{Distribution of the field normalized energy density $W(\protect\xi ,0,\protect\zeta ,\protect\tau )$ in the CNT array, at various moments of dimensionless time $%
\protect\tau =\protect\omega _{0}t/\protect\sqrt{\protect\varepsilon }$, in
the case when the incident laser pulse is reflected from the HED
(high-electron-density) layer, placed at $\protect\zeta =0$, in the absence
of the control (external) HF field [$\protect\kappa =0$, see Eq. (\protect
\ref{eq:kappa})]: (a) $\protect\tau =0$; (b) $\protect\tau =3.0$; (c) $\protect%
\tau =6.0$; (d) $\protect\tau =9.0$. The dimensionless coordinates are $%
\protect\xi =x\protect\omega _{0}/c$ and $\protect\zeta =z\protect\omega %
_{0}/c$ [see Eq. (\protect\ref{tau-xi})]. The color code shows values of the energy density normalized  as $W/W_{0}$ [see Eq. (\protect\ref{I0})], the
yellow and blue areas corresponding to the maximum and minimum values of the field energy density, respectively.}
\label{fig2}
\end{figure}

\begin{figure}[tbp]
\includegraphics[width=1.0\textwidth]{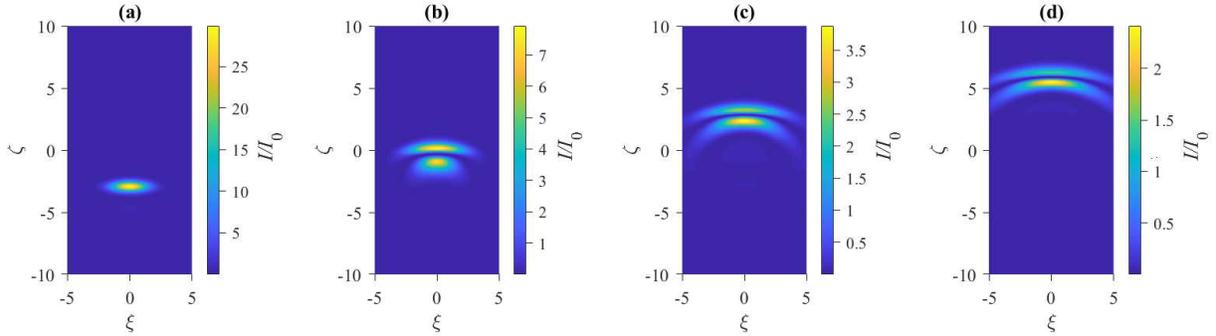}
\caption{The same as in \protect\ref{fig2}, but for the case when the
incident pulse passes the HED layer, in the presence of the control HF
field, with strength $\protect\kappa =2.0$ [see Eq. (\protect\ref{eq:kappa})]:
(a) $\protect\tau =0$; (b) $\protect\tau =3.0$; (c) $\protect\tau =6.0$; (d)
$\protect\tau =9.0$. }
\label{fig3}
\end{figure}

Figures \ref{fig2} and \ref{fig3} display the interaction of the ultrashort
electromagnetic pulse with the HED layer for various values of strength $\kappa$ (see Eq.~\eqref{eq:kappa}) of the control HF field and fixed values of the initial pulse's parameters $%
\Omega =0.5$ and $w_{0}=2.0$. In this case, the maximum value of the
electric field of the pulse is $|E_{x}|_{\text{\text{max}}}\approx 6.1\times
10^{5}$ V/cm [see Eq.~\eqref{19}], and its duration is $\Delta t_{\text{pulse%
}}\approx 0.67\ $ps, see Eq.~\eqref{20}. For the
definiteness, we have selected the following fixed values of parameters of
the HED\ layer: $\eta _{\text{imp}}^{\text{max}}=30$ and $\delta \zeta _{%
\text{imp}}=0.1$.

Figures \ref{fig2} and \ref{fig3} display the distribution of the field energy density of the ultrashort pulse, $W(\xi ,0,\zeta ,\tau )$, see Eq.~%
\eqref{11}, in the $\left( \xi ,\zeta \right) $ plane (at $\upsilon =0$) at
various moments of the dimensionless time, $\tau =\omega _{0}t/\sqrt{%
\varepsilon }$. The normalized field energy density is represented by
code-colored values of $W/W_{0}$, with $W_{0}$ defined as per Eq.~\eqref{I0}. The horizontal and vertical dimensionless coordinates are $\xi =x\omega
_{0}/c$ and $\zeta =z\omega _{0}/c$, which are defined above in Eq.~\eqref
{tau-xi}. With the values of the system's parameters selected above, $\xi
=1 $ and $\zeta =1$ correspond to distance $\approx 4.2\times 10^{-3}$ cm.
Note that in these figures we present the distribution of the energy density only
in the $\left( \xi ,\zeta \right) $ plane (at $\upsilon =0$), as it is
identical to that in the $\left( \upsilon ,\zeta \right) $-plane.

Figure \ref{fig2} shows an example of the reflection of the ultashort
electromagnetic pulse from the HED layer in the absence of the control HF
field ($\kappa =0$). The initial pulse's speed is taken as $U=0.8$, which
corresponds to the initial speed $u=0.012$ cm/ps in physical units. In this
case, the transmission and reflection coefficients, defined by Eqs.~\eqref{12} and~\eqref{13}, are found to be $K_{\text{pass}}\approx 0.0143$ and $K_{%
\text{refl}}\approx 0.9857$, respectively, confirming the nearly complete
reflection of the incident pulse (small-amplitude transmitting waves are not
visible in Fig. \ref{fig2}).

Figure \ref{fig3} shows the opposite situation,\textit{\ viz}., the passage
of the pulse, with the same parameters as in Fig. \ref{fig2}., through the
same HED layer, in the presence of the control HF field with amplitude $%
E_{10}=5.8\times 10^{6}$ V/cm and frequency $\omega _{1}=9.4\times 10^{13}$ s%
$^{-1}$, which corresponds to $\kappa \approx 2.0$, as per Eq.~\eqref{eq:kappa}. In this case, the transmission and reflection coefficients are $K_{\text{%
pass}}\approx 0.9196$ and $K_{\text{refl}}\approx 0.0804$, respectively,
which confirms the practically complete passage, with a small reflected wave
being virtually invisible in Fig. \ref{fig3}. Note that, at $\kappa \approx
2.0$, significantly stronger diffraction spreading of the pulse is observed,
in comparison to the case shown in Fig. \ref{fig2}, as in present case the
pulse propagates in the medium with properties of the electronic subsystem
dynamically modified by the control HF field, therefore self-focusing
properties of this medium are less pronounced in comparison with the case of
$\kappa =0$.

\begin{figure}[tbp]
\includegraphics[width=1.0\textwidth]{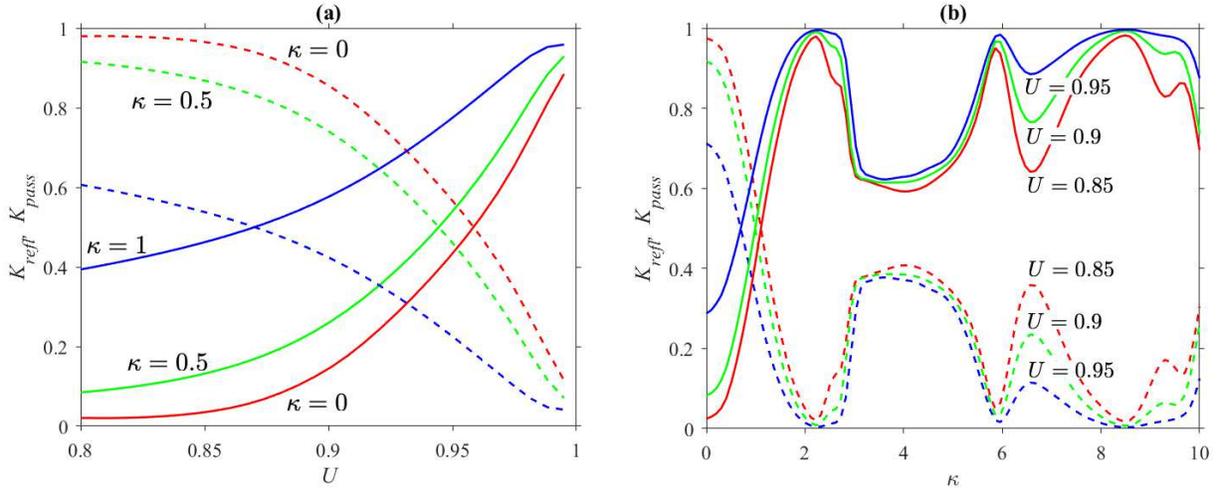}
\caption{(a) The transmission and reflection coefficients, $K_{\text{pass}}$
and $K_{\text{refl}}$ (the solid and dashed lines, respectively) as
functions of the initial speed $U$ of the incident electromagnetic pulse,
for fixed values of strength $\protect\kappa $ of the control field attached
to each curve. (b) The same coefficients as functions of $\protect\kappa $,
at fixed values of $U$ attached to the curves. In both panels, marked are
threshold values $U_{\text{thr}}$ and $\protect\kappa _{\text{thr}}$, at
which the transmission and reflection coefficients are equal.}
\label{fig4}
\end{figure}

Figure \ref{fig4} shows dependencies of the reflection and transmission
coefficients $K_{\text{pass}}$ and $K_{\text{refl}}$ on initial speed $U$,
longitudinal width and duration (for a fixed value of strength $\kappa $ of
the control HF field), and on $\kappa $ (for a fixed value of $U$). As can
be seen in Fig.~\ref{fig4}(a), the transmission and reflection coefficients
increase and decrease, respectively, with the increase of velocity $U$, cf.
Ref.~\cite{25}. When the initial velocity of the incident ultrashort pulse
significantly exceeds a certain threshold value, $U\gg U_{\text{thr}}$, the
passage of the pulse through the HED layer prevails over the reflection. At $%
U=U_{\text{thr}}$ the transmission and reflection coefficients are equal, $%
K_{\text{refl}}(U_{\text{thr}})\approx K_{\text{pass}}(U_{\text{thr}})=1/2$,
with the incident pulse splitting in two approximately identical wave
packets, one of which continues to move in the original direction, while the
other one bounces back. Figure \ref{fig4}(a) shows that, with the increase
of strength $\kappa $ of the control HF field, the threshold velocity $U_{%
\text{thr}}$ shifts to lower values. In other words, the control field
uphold the passage of the short pulse through the HED layer.

In Fig. \ref{fig4}(b), the coefficients become equal, $K_{\text{pass}}=K_{%
\text{refl}}$, at strength $\kappa $ of the control field equal to the
respective threshold value, $\kappa _{\text{thr}}$. A noteworthy result,
clearly seen in the figure, is that $K_{\text{pass}}$ and $K_{\text{refl}}$
vary non-monotonously as functions of $\kappa $ at $\kappa >\kappa _{\text{%
thr}}$. In particular, the reflection coefficient features almost zero
values at its local minima, which may be considered as \textit{transparency
windows} of the medium for the propagation of the ultrashort electromagnetic
pulses. 

Using Eq. ( \ref{8}), and taking into account the fact that, in a rough approximation,
 all $G_r$ with $r \neq 1$ can be neglected, 
the current in the absence of HF field  can be expressed as 
\begin{equation}
\langle j \rangle\approx-en\frac{d_{x}}{\hbar }\gamma _{0}J_{0}\left( r\frac{%
|eE_{10}|d_{x}}{\hbar \omega _{1}}\right) G_{1}\sin \left( rA\frac{ed_{x}}{%
c\hbar }\right) ,  \label{8bis}
\end{equation}
i.e.,  due to the electronic properties of the CNTs, it expresses as
 a periodic function of the vector potential.
 The external HF field acts as a phase shift in this periodic function. Only the mean value of the current over the fast oscillations has an effect on the infrared pulse propagation but, due to this dependence, this mean value vanishes for specific values of the HF field intensity only
(unless either the the infrared pulse or the HF field strongly dominates, of course).
These values correspond to specific value of the parameter $\kappa$, 
which can be seen as the ratio of two energies, say $\kappa=W_{CNT}/W_{HF}$, where $W_{CNT}=E_{10}e d_x$ is the typical magnitude of the energy of the electric dipolar momentum of the CNT in the HF field,  and $W_{HF}=\hbar\omega_1$ the energy of the HF field photon. 
The transparency windows appear thus as a resonant effect, which occurs when the photon energy  matches the energy of the CNT dipolar momentum. 

This effect resembles
the phenomenon of the electromagnetically-induced transparency (EIT)~\cite{EIT0,EIT,EIT2,EIT3,EIT4,EIT5,destructive}, where the photonic resonance is the key factor. The main difference between the problem under consideration and the problem with EIT is that the EIT effect is observed in an environment in which an electromagnetic field propagates in a medium, considered as a set of discrete well-separated energy levels of atoms. In our case, the electromagnetic field propagates within the medium of electrons located in the conduction band (or holes in the valence band) of the carbon nanotubes. The spectrum in our case is continuous. We also note that the EIT effect leads to a deceleration of the pulse of the electromagnetic field. This is achieved by saturating the population at certain levels (depending on the level scheme). In the present study, given the continuity of the spectrum, the concept of saturation is irrelevant, and accordingly, the pulse does not slow down as in the EIT problems.

The location of the windows on the scale of $\kappa $
may be explained by rewriting current \eqref{8} in terms of $\kappa $, taking into account that, in a rough approximation, all $G_r$ with $r \neq 1$ can be neglected, and according to Eq.~\eqref{eq:kappa}:%
\begin{equation}
\langle j\rangle \approx -en\frac{d_{x}}{\hbar }\gamma _{0}J_{0}(\kappa )G_1\sin \left( A%
\frac{ed_{x}}{c\hbar }\right) .  \label{j}
\end{equation}%
Indeed, it is easy to see that bottom points of the windows in Fig. \ref%
{fig4}(b) are relatively close to zeros of function $J_{0}(\kappa )$ in Eq.~\eqref{j}. Actually, the Bessel factor $J_{0}(\kappa )$
represents the result of the above-mentioned resonance effect. 

Because the nonlinearity nearly vanishes in Eq.~\eqref{10} in the region of
the transparency windows, dynamics of the ultrashort pulse becomes nearly
linear in such cases. This fact explains the spatiotemporal evolution in
Fig.~\ref{fig3}, in the course of the passage through the HED layer, the
pulse undergoes noticeable transverse and longitudinal deformation due to
diffraction and dispersion, which is characteristic for the propagation in
quasi-linear media. On the contrary, the deformation is not prominent in
Fig. \ref{fig2}, which pertains to the case of strong nonlinearity.

Thus, the result of the interaction of the short laser pulse with the HED
layer depends on the values of several parameters, among which a special
role is played by the amplitude and frequency of the control HF field, which
may be used as the most effective control parameters of the system, as they
may be varied without changing the sample's structure and/or standard pulses
employed by the scheme. In particular, these parameters may be efficiently
used to change, as required, the threshold velocity of the incident pulse, $%
U_{\text{thr}}$, which separates the reflection and passage outcomes of the
collision of the pulse with the HED barrier. 
\begin{figure}[htbp]
\includegraphics[width=1.0\textwidth]{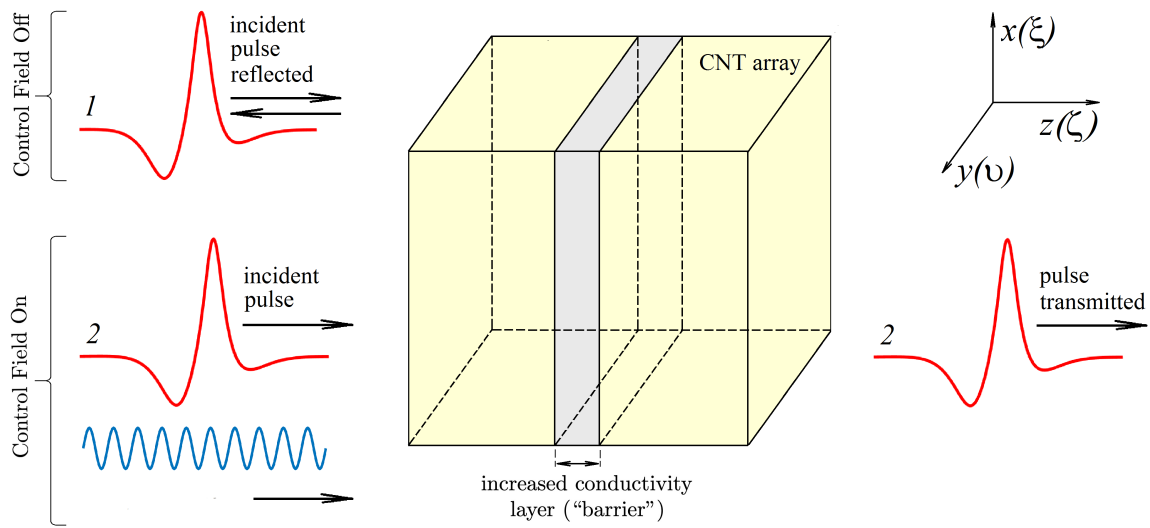}
\caption{The schematic of a \textquotedblleft soliton valve", based on the
HED layer embedded in the CNT\ array. (\textit{1}) In the absence of the
control HF field, the incident pulse is reflected fby the barrier layer. (%
\textit{2}) The application of the control field unlocks the barrier,
letting the pulse pass through it.}
\label{fig5}
\end{figure}

A conclusion is that the scheme elaborated here may be used as a
\textquotedblleft soliton valve", somewhat similar to the concept of
all-optical transistors (\textquotedblleft light controlled by light") \cite%
{transistor1,transistor2}. An illustration is presented in Fig. \ref{fig5}:
While, in the absence of the control HF field, the HED layer does not let
the incident short pulse pass the barrier, one can open the passage by
applying the control field with appropriate values of the amplitude and
frequency. 


\section{Conclusions}


This study reports four key results, which have some far-reaching practical applications in the area of design and development of optoelectronic products and systems.

First, we establish that, as a result of the interaction of the
ultrashort electromagnetic pulse with the barrier layer of HED (high
electron density) in an array of CNTs (carbon nanotubes), the pulse can
either pass through the layer or be reflected from it. This first result is important as it creates the possibility to effectively use variable HED layers to achieve some specific propagation or reflection of ultra-short pulses.

Second, our analysis of this rich phenomenon reveals that the collision of the pulse with the HED layer has a very particular dependency on both the parameters of the pulse and the properties of the HED layer itself. Specifically, we found that an increase of the speed and amplitude of the incident pulse facilitates the passage of the pulse through this layer. This property paves the way to a particular modulation of the pulse through its speed and amplitude in order to control its penetration effectiveness.

Third, we establish the central result of this study, which is built upon the two previous key results. It has been found that the state of the electronic subsystem in the CNT array may be controlled by an external HF field. That control field may be used to facilitate the passage of the ultrashort pulse through the barrier layer, thereby creating transparency windows around particular values of the control-field's amplitude. This effect, based in the resonant effect of the HF fields, is similar to the phenomenon of the electromagnetically-induced transparency.

Lastly, and from the practical standpoint, we proved that the latter effect can be used for the design of a soliton valve, which allows to efficiently switch the system between the reflection and transmission regimes for the soliton stream, without changing its parameters, but rather adjusting the amplitude and frequency of the control field.

\begin{acknowledgments}
N. N. Rosanov acknowledges support from the Russian Foundation for Fundamental Investigations (Grant 16-02-00762) and from the Foundation for the Support of Leading Universities of the Russian Federation (Grant 074-U01). N. N. Konobeeva and M. B. Belonenko acknowledges support from the Ministry of Science and Higher Education of the Russian Federation for the numerical modeling and parallel computations support under the government task (no. 0633-2020-0003)
\end{acknowledgments}

\appendix
\section{Detailed derivation of Equation~\eqref{8}}
This appendix provides a detailed derivation of Eq.~\eqref{8}:
\begin{equation*}
j=-en\frac{d_{x}}{\hbar }\gamma _{0}\sum_{r=1}^{\infty }J_{0}\left( r\frac{%
|eE_{10}|d_{x}}{\hbar \omega _{1}}\right) G_{r}\sin \left( rA\frac{ed_{x}}{%
c\hbar }\right) .
\end{equation*}%
As a starting point, we replace $A$ by $(A+A_1)$ inside the term $\sin \left( rA\frac{ed_{x}}{c\hbar }\right)$ in Eq.~\eqref{4}. As specified by Eq.~\eqref{7}, this amounts to replacing $A$ by $A-E_{10}\frac{c}{\omega _{1}}\sin (\omega _{1}t+\alpha )$ in the rightmost sine term in Eq.~\eqref{8}:
\begin{equation}
\sin \left( rA\frac{ed_{x}}{c\hbar }\right)
\rightarrow 
\sin \left( r\left( A-E_{10}\frac{c}{\omega _1}\sin (\omega _{1}t+\alpha)\right)\frac{ed_{x}}{c\hbar }\right),
\end{equation}
with
\begin{eqnarray}
\sin \left( r\left( A-E_{10}\frac{c}{\omega _1}\sin (\omega _{1}t+\alpha)\right)\frac{ed_{x}}{c\hbar }\right)&=\sin \left( rA\frac{ed_{x}}{c\hbar }\right) \cos \left( \rho \sin \beta \right)\nonumber \\
-&\cos \left( rA\frac{ed_{x}}{c\hbar }\right) \sin \left( \rho \sin \beta\right),\label{A2}
\end{eqnarray}
where the following two quantities have been introduced to simplify the notations:
\begin{align}
\rho & = rE_{10}\frac{c}{\omega _1}\frac{ed_{x}}{c\hbar },\\
\beta & = \omega_1 t + \alpha.
\end{align}
Using Bessel functions of the first kind, one can obtain that
\begin{align}
\cos \left( \rho \sin \beta \right) & = J_0(\rho)+2\sum_{k=1}^{\infty}J_{2k} (\rho)\cos (2k\beta),\\
\sin \left( \rho \sin \beta   \right) & =2 \sum_{k=1}^\infty J_{2k-1}(\rho) \sin \left( (2k-1)\beta \right),
\end{align}
where $J_\mu$ is the $\mu$-th order Bessel function of the first kind~\cite{31}. Substituting these last two equations into Eq.~\eqref{A2}, and subsequently in Eq.~\eqref{4}, we obtain the expression for the electric current density
\begin{eqnarray}
j=-en\frac{d_{x}}{\hbar }\gamma _{0}\left\{
\sum_{r=1}^{\infty } G_r \sin \left( rA\frac{ed_{x}}{c\hbar }\right) 
J_0 \left(  rE_{10}\frac{c}{\omega _1}\frac{ed_{x}}{c\hbar }\right)\right. \nonumber\\
+2\sum_{r=1}^{\infty }\sum_{k=1}^\infty G_r \sin \left( rA\frac{ed_{x}}{c\hbar }\right) J_{2k} \left(rE_{10}\frac{c}{\omega _1}\frac{ed_{x}}{c\hbar }\right)\cos (2k(\omega_1 t + \alpha))\nonumber \\
\left.+2\sum_{r=1}^{\infty }\sum_{k=1}^\infty G_r \cos \left( rA\frac{ed_{x}}{c\hbar }\right) J_{2k-1} \left(rE_{10}\frac{c}{\omega _1}\frac{ed_{x}}{c\hbar }\right)\sin ((2k-1)(\omega_1 t + \alpha))
\right\}.\label{A7}
\end{eqnarray}
As a next step, we average the electric current density $j$ in Eq.~\eqref{A7} over the period $T_1=2\pi/\omega_1$ of the high-frequency external electric field and obtain the desired equation:
\begin{equation}
\langle j\rangle=-en\frac{d_{x}}{\hbar }\gamma _{0}\sum_{r=1}^{\infty }J_{0}\left( r\frac{%
|eE_{10}|d_{x}}{\hbar \omega _{1}}\right) G_{r}\sin \left( rA\frac{ed_{x}}{%
c\hbar }\right) .
\end{equation}

\end{document}